\documentstyle[epsfig,11pt]{article}
\newcommand{\be}{\begin{equation}}
\newcommand{\ee}{\end{equation}}
\newcommand{\ben}{\begin{eqnarray}}
\newcommand{\een}{\end{eqnarray}}
\newcommand{\bc}{\begin{center}}
\newcommand{\ec}{\end{center}}

\begin{document}

\title{Chromaticity effects in microlensing by wormholes}
\author{Ernesto Eiroa$^{1,}$\thanks{e-mail: eiroa@iafe.uba.ar}, Gustavo E.
Romero$^{2,}$\thanks{Member of CONICET; e-mail:
romero@irma.iar.unlp.edu.ar }, and Diego F.
Torres$^{2,}$\thanks{e-mail: dtorres@venus.fisica.unlp.edu.ar}
\\{\small $^1$ Instituto de Astronom\'{\i}a y F\'{\i}sica del
Espacio, C.C. 67, Suc. 28, 1428, Buenos Aires, Argentina}\\
{\small $^2$Instituto Argentino de Radioastronom\'{\i}a, C.C.5, 1894
Villa Elisa, Buenos Aires, Argentina}  }

\maketitle

\begin{abstract}
Chromaticity effects introduced by the finite source size in
microlensing events by presumed natural wormholes are studied. It
is shown that these effects provide a specific signature that
allow to discriminate between ordinary and negative mass lenses
through the spectral analysis of the microlensing events. Both
galactic and extragalactic situations are discussed.

\end{abstract}

\newpage

\section{Introduction}

A wormhole is a region of space-time with non-trivial topology. It
has two mouths connected by a throat. The mouths are not hidden by
event horizons, as in the case of black holes, and, in addition,
there is no singularity that could avoid the passage of particles
from one side to the other. After the pioneering paper by Morris
\& Thorne \cite{motho}, wormhole solutions to the Einstein field
equations has been extensively studied in the literature (see
Refs. \cite{wh} and references cited therein).\\

Static wormhole structures require the violation of the average
null energy condition in the wormhole throat in order to exist.
Plainly stated, this means that the matter threading the wormhole
must exert gravitational repulsion to stay stable against
collapse. Although there are known violations to the energy
conditions (e.g. the Casimir effect), it is currently far from
clear whether large macroscopic amounts of ``exotic matter'' exist
in the nature. If natural wormholes actually exist in the universe
(e.g. if the original topology after the Big-Bang was multiply
connected), then there should be observable electromagnetic
signatures that could lead to their identification.\\

Cramer et al. \cite{cramer} and Torres et al. \cite{torr} have
recently studied the microlensing effects of a negative mass (e.g.
a natural stellar-size wormhole) on the light from a background
point-like source [for a macrolensing study see \cite{S}]. These
authors have shown that the typical lightcurves expected from
microlensing events produced by wormholes should be very different
from the lightcurves in ordinary (positive mass) microlensing of
point sources. In the standard case, a time symmetric burst in the
flux density from the background source occurs. If the lens,
instead, is a wormhole, the gravitational repulsion creates an
obscure umbra region, deflecting light rays  that should,
otherwise, reach the observer. The wormhole basically acts as a
divergent lens. However, light is concentrated on the border of
the umbra region producing two separated intensity enhancements
that are observed before and after the occultation event. These
two burst are individually asymmetric under time reversal,
although they are mirror images of each other. In the first burst,
the observed flux increases towards a divergency and then drops to
zero. The infinity, or ``caustic", and the vertical drop to zero
are consequences of the ideal model adopted in the calculations,
with a point source at the background. A more realistic model
should consider an extended
source with a given intensity distribution over it.\\

Recently, Anchordoqui et al. \cite{doqui} searched in existent
gamma-ray bursts databases for signatures of wormhole
microlensing. Although they detected some interesting candidates,
no conclusive results could be obtained. Peculiarly asymmetric
gamma-ray bursts \cite{rom}, although highly uncommon, might be
probably explained by more conventional hypothesis, like
precessing fireballs (see, for instance, Ref. \cite{Zwart}). Even
in the case of galactic microlensing, highly asymmetric curves can
be the effect of microlensing by binary stellar systems
\cite{bin}. Hence, although binary lenses produce asymmetric
lightcurves with different shapes than those that would be
produced by wormholes, discriminators other than the lightcurves
are needed in order to identify wormhole microlensing events.\\

In this paper, we expand the analysis of Cramer et al.
\cite{cramer} and Torres et al. \cite{torr} to the extended source
case. This allow us to present more realistic lightcurves for
wormhole microlensing events. Then, using this formalism, we
compute the effects of a finite source extent on the spectral
features of microlensing. We show that limb darkening of the
intensity distribution on a stellar source induces specific
chromaticity effects that are very different from what is expected
in the positive mass lens case. Thus, multi-color optical
observations can be used to search for galactic natural wormholes.
In addition, we study the extragalactic wormhole microlensing of
background AGNs taking into account the variations of the source
size with the observing wavelengths. This extends the previous
work by Torres et al. \cite{torr}, who first considered these kind
of events at gamma-rays adopting a point-like source. We show that
also in this case the spectral evolution during the wormhole
microlensing event presents peculiar signatures that allow its
identification.\\

\section{Microlensing by wormholes in the extended source case}

The total amplification $A=I_{\rm obs}/I_0$ --where $I_{\rm obs}
(I_0)$ is the observed (intrinsic) luminosity-- of a background
point source due to gravitational lensing by a single point mass
is given by \cite{cramer}
\begin{equation}
A_{0}=\frac{B_{0}^{2}\pm 2}{B_{0}\sqrt{B_{0}^{2}\pm 4}},
\label{1}
\end{equation}
where the plus (minus) sign corresponds to the positive (negative)
mass case, and
\begin{equation}
B_{0}=\frac{b_{0}}{R_{\rm E}}  \label{2}
\end{equation}
is the shortest lens-source separation (on the lens plane) in
units of the Einstein ring radius, $R_{\rm E}$, defined as
\begin{equation}
R_{\rm E}=\sqrt{\frac{4G|M|}{c^{2}}\frac{D_{\rm ol}D_{\rm
ls}}{D_{\rm os}}}. \label{3}
\end{equation}
In this last expression,  $D_{\rm os}$ is the observer-source
distance, $D_{\rm ol}$ is the observer-lens distance, $D_{\rm ls}$
is the lens-source distance, and $M$ is the mass of the
gravitational lens.\footnote{It should be kept in mind that any
form of negative masses, not only wormholes,
would produce the same observational effects.}\\

When the source is extended, we must take into account the
contributions coming from the different parts of it. The
amplification observed at a frequency $\nu$, then, results:
\begin{equation}
A_{\nu }=\frac{\int_{0}^{2\pi}\int_{0}^{r_{*}}{\mathcal I}_{\nu
}(r,\varphi )A_{0}(r,\varphi )rdrd\varphi
}{\int_{0}^{2\pi}\int_{0}^{r_*}{\mathcal I}_{\nu }(r,\varphi
)rdrd\varphi },  \label{4}
\end{equation}
where $(r,\varphi )$ are polar coordinates in a reference frame
centered in the source (e.g. a star), $r_{*}$ is the radius of the
source and ${\mathcal I}_{\nu }( r ,\varphi )$ is its surface
intensity distribution. For a radially symmetric distribution we
have:
\begin{equation}
A_{\nu }=\frac{\int_{0}^{2\pi}\int_{0}^{r_{*}}{\mathcal I}_{\nu
}(r)A_{0}(r,\varphi )rdrd\varphi }{2\pi \int_{0}^{r_*}{\mathcal
I}_{\nu }(r)rdr}. \label{5}
\end{equation}
If the lens is moving with constant velocity $v$, the lens-source
separation evolves in time as
\begin{equation}
b(t)=\sqrt{\left( b_{0}+r\sin \varphi \right) ^{2}+\left(
-vt+r\cos \varphi \right) ^{2}}.  \label{11}
\end{equation}
In the published version of this paper, this is illustrated in the
first figure. 
In units of the Einstein radius, we have:
\begin{equation}
B(T)=\frac{b(t)}{R_{E}}=\sqrt{\left( B_{0}+R\sin \varphi \right)
^{2}+\left(-T+R\cos \varphi \right) ^{2}},  \label{12}
\end{equation}
where $ T=vt/R_{E}$. Replacing now $B_{0}$ by $B(T)$ in Eq.
(\ref{1}), we obtain the time dependent amplification.\\

In order to compute the expected lightcurves for the case of
negative mass lenses, we take the negative sign in Eq. (\ref{1})
and replace it in Eq. (\ref{5}). In Figure 2 we show different
lightcurves for different impact parameters, calculated assuming a
circular source of uniform brightness and radius
$R_*=r_*/R_E=0.1$. These curves should be compared with the ideal
curves presented by Torres et al. \cite{torr} in their Figure 1.
The main difference is that in the extended source case the
divergencies are eliminated, and there is not a discontinuous
transition to or from the umbra region. The basic features of the
microlensing event, however, remain. For $B_0<2$, we have two
successive intensity enhancements, separated by a period of
absence of radiation. The individual bursts are quite asymmetric.
In the case $B_0>2$, there is only one event, symmetric under time
reversal and similar to what is observed in the standard case.
\begin{figure}[t]
\begin{center}
\includegraphics[width=8cm,height=7cm]{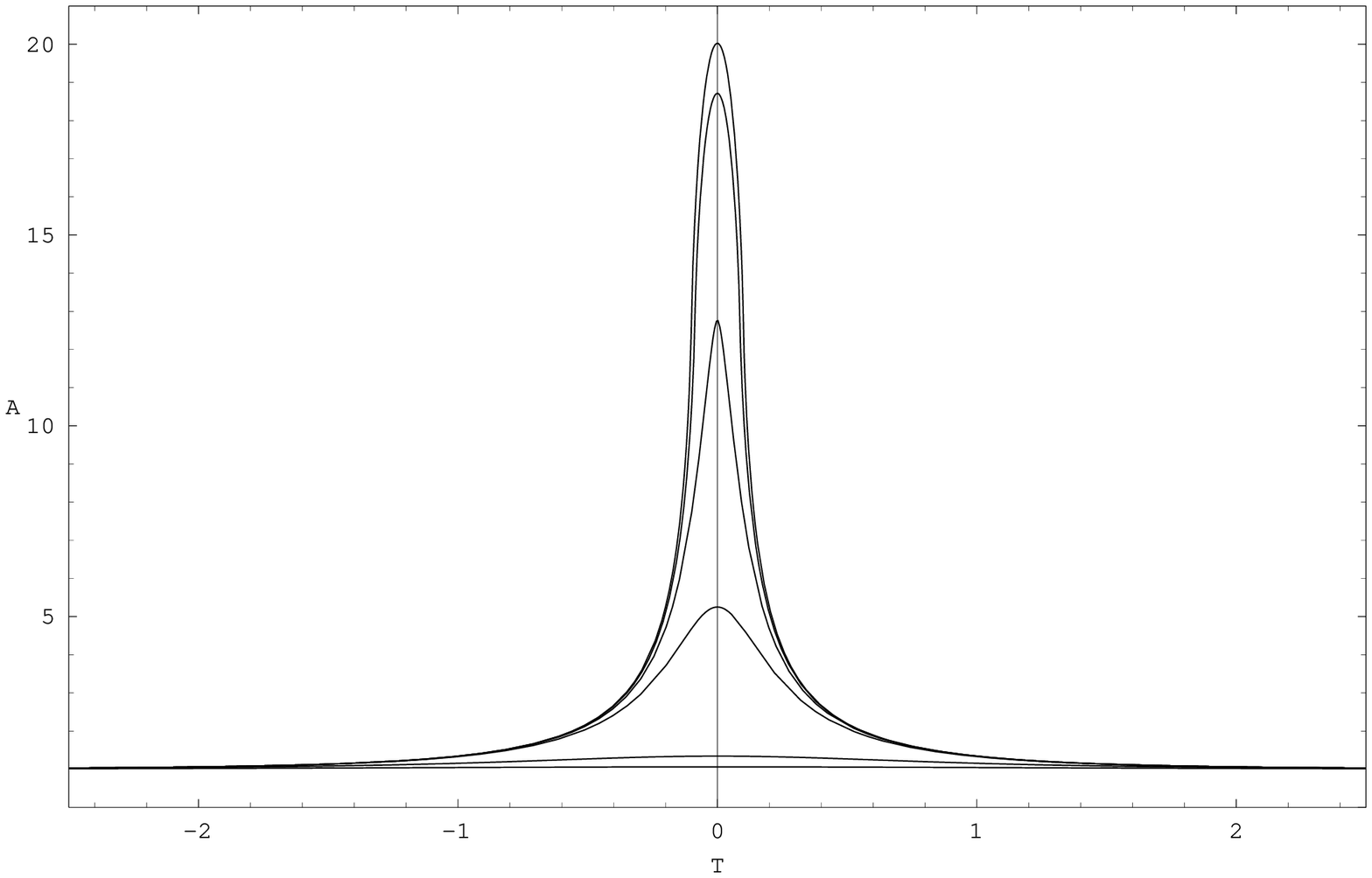}%
\includegraphics[width=8cm,height=7cm]{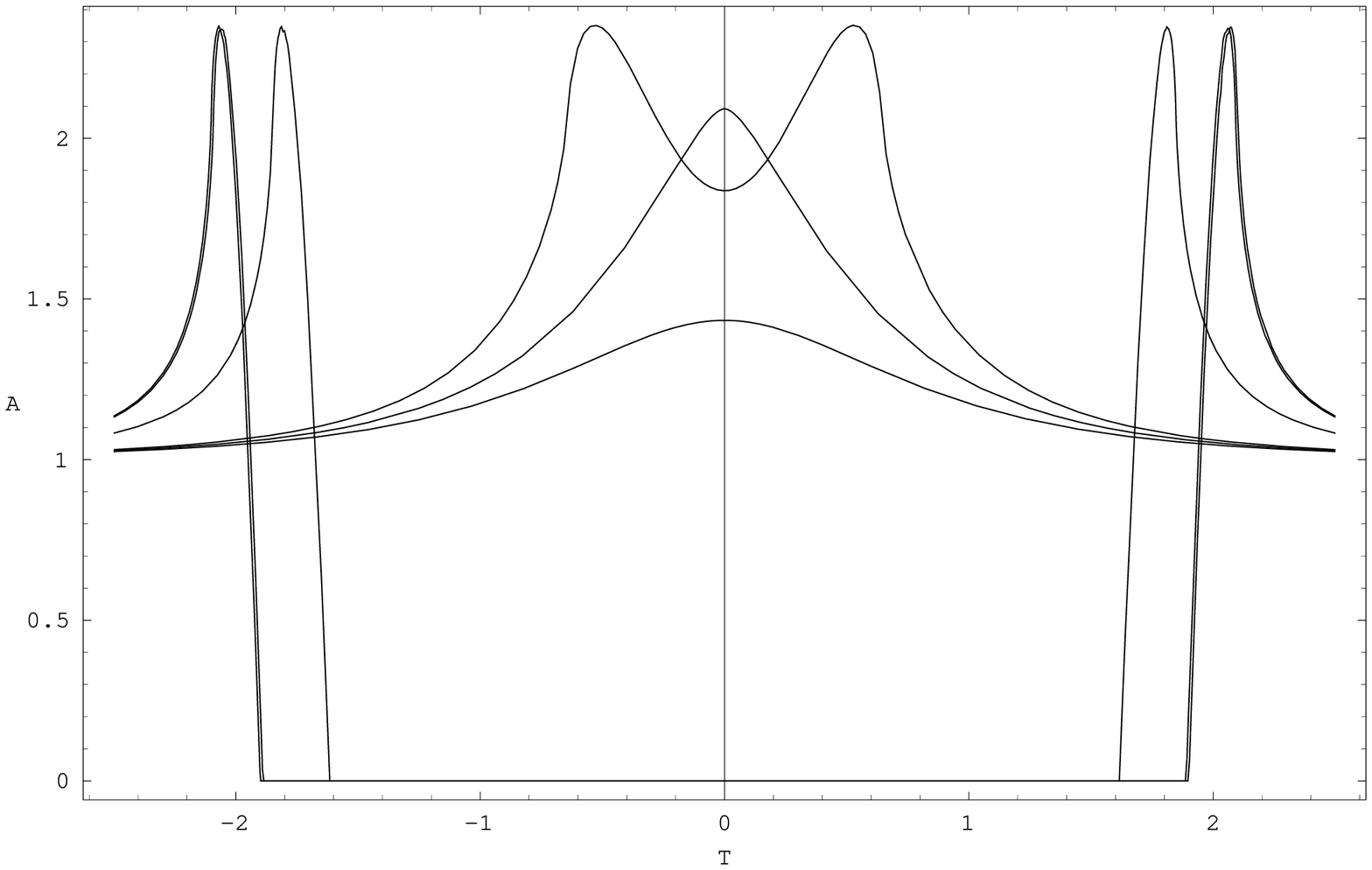}
\end{center}
\vspace{-3cm} \caption{Lightcurves for a microlensing event of a
background extended source produced by a positive point mass (left
panel) and a negative point mass (right panel). The vertical axis
gives the total amplification of the background intensity. The
different curves are for different impact parameters [$k$=0, 0.5,
1, 2, 10, and 20, from top to bottom, in the case of positive
lensing; and $k$=0 and 2 (in the same curve), 10, 20, 21, and 22,
from left to right, in the case of negative lensing]. See text for
additional details. \label{fig1}}
\end{figure}

Real background sources in microlensing events can present
non-uniform brightness distributions on their surfaces and a
dependency of their emission with the observing frequency. These
complications can result in chromaticity effects, i.e. in spectral
changes induced by differential lensing during the event. In the
next sections we shall compute such changes in order to establish
whether they can provide a specific signature of wormhole
microlensing.\\

\section{Chromaticity effects in microlensing of stars}

Stars are brighter in their center. The obscuration of the
intensity profile of a star towards its border is known as ``limb
darkening".  They also radiate as a blackbody with an effective
temperature $T_{\rm eff}$, and consequently their emission is
frequency dependent. To compute the chromaticity effects due to
these characteristics, we represent the intensity profile of
typical star by \cite{Han}:
\begin{equation}
{\mathcal I}_{\nu }(r)=1-C_{\nu }\left( 1-\sqrt{1-\left(
\frac{r}{r_{*}} \right) ^{2}}\right),  \label{6}
\end{equation}
where the limb-darkening coefficients in the I and U bands can be
taken as $C_{\nu _{1}}=0.503$ and $C_{\nu _{2}}=1.050$,
respectively, for a K-giant star with $T_{\rm eff}=4750$K.\\

Defining the dimensionless radius as $R=r/R_{E}$, we obtain
\begin{equation}
A_{\nu }=\frac{\int_{0}^{2\pi}\int_{0}^{R_{*}}{\mathcal I}_{\nu
}(R)A_{0}(R,\varphi )RdRd\varphi }{2\pi \int_{0}^{R_*} {\mathcal
I}_{\nu }(R)RdR} \label{8}
\end{equation}
and
\begin{equation}
{\mathcal I}_{\nu }(R)=1-C_{\nu }\left( 1-\sqrt{1-\left(
\frac{R}{R_{*}} \right) ^{2}}\right),  \label{9}
\end{equation}
where $ R_{*}=r_{*}/R_{E}$ is the dimensionless radius of the
star. The colour change caused by limb darkening can then be
computed by \cite{Han}
\begin{equation}
\Delta (m_{\nu _{1}}-m_{\nu _{2}})=-2.5\log \left( \frac{A_{\nu
_{1}}}{ A_{\nu _{2}}}\right).  \label{15}
\end{equation}

\begin{figure}[t]
\begin{center}
\includegraphics[width=8cm,height=7cm]{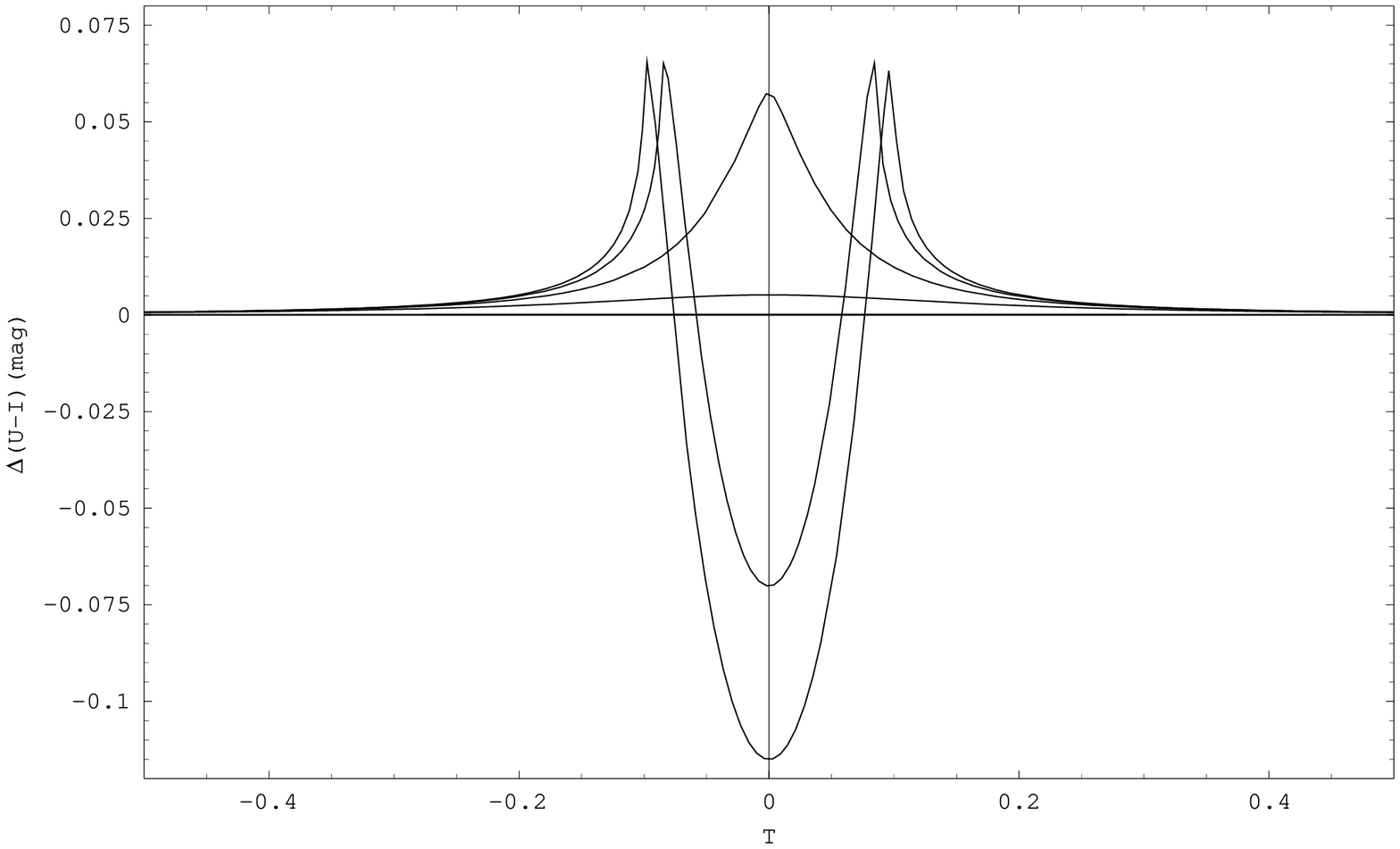}%
\includegraphics[width=8cm,height=7cm]{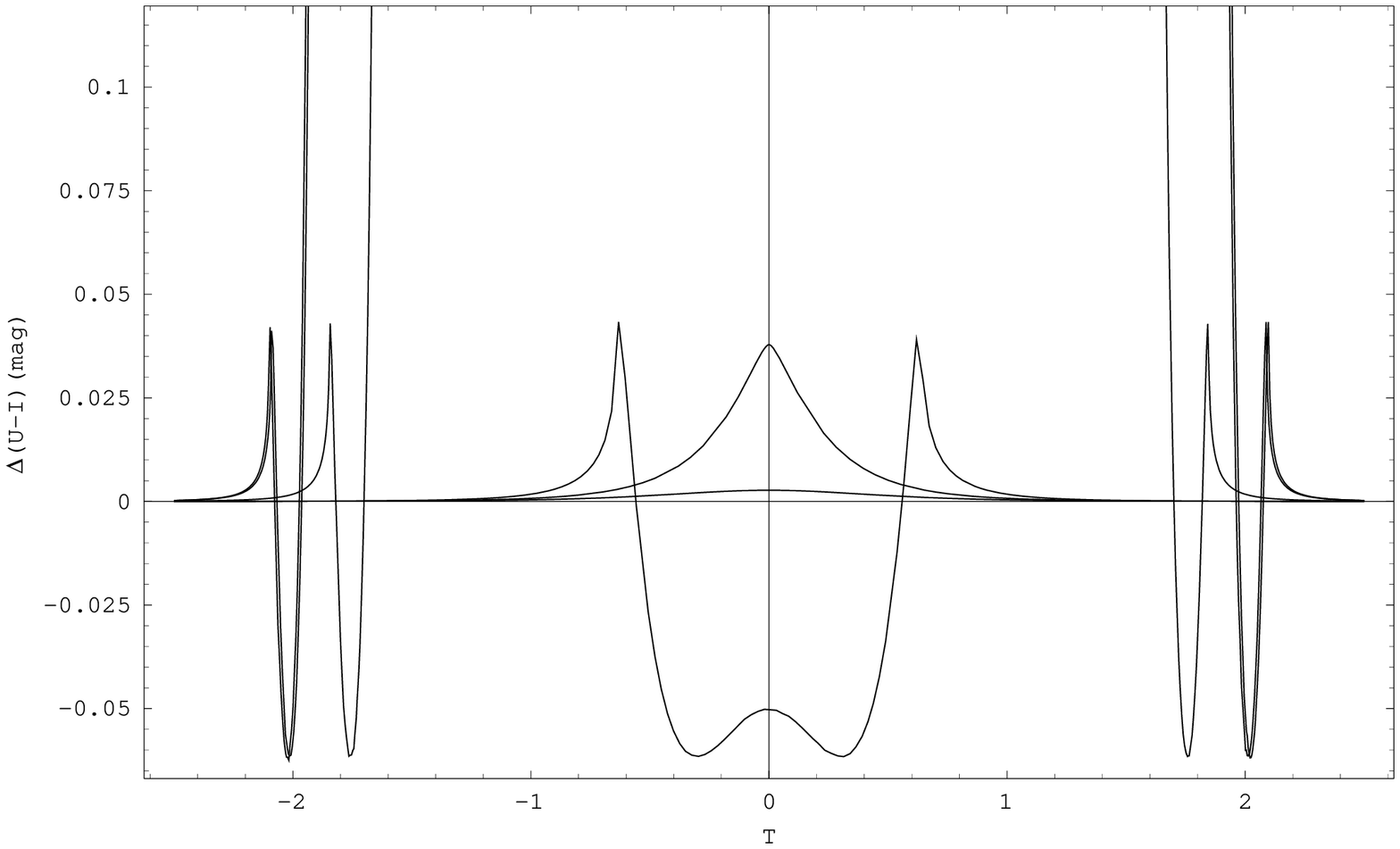}
\end{center}
\vspace{-3cm} \caption{Color curves for a microlensing event of a
background extended source of variable surface brightness produced
by a positive point mass (left panel) and a negative point mass
(right panel). The different curves are for the same impact
parameters as in Figure 2. In the case of negative mass lensing,
from left to the right, $k$= 0 and 2 (in the same curve), 10, 20,
21, and 22. In the case of the positive mass lensing, from left to
right, $k$=0, 0.5, 1, 2, 10 and 20 (the latter two in the same
curve). See text for additional details. \label{fig2}}
\end{figure}
In Figure 3 we show several color curves for both positive (left
panel) and negative (right panel) mass lenses. The different
curves correspond to different impact parameters $k=B_{0}/R_{*}
\label{e11}$ specified in the caption. The star radius was
considered, as before, equal to $R_*=0.1$. In the case of an
ordinary lens, we see that as it gets closer to the star, the
color of the observed source becomes redder due to the
differential amplification of the coldest regions. When the lens
transits towards the star interior, the hot center starts to
dominate the amplification, producing a dramatic change in the
slope of the color curve. This is basically the result recently
found by Han et
al. \cite{Han}.\\

If we now look at the color curves for the negative mass lens
case, we notice a clear contrast in the behaviour. The spectral
changes start long before than in the standard situation.
Initially, the source also becomes redder and then experiences a
switch when shorter wavelengths begin to dominate. Contrary to
what happens with positive masses, the spectral trend changes
again, with the source appearing colder and colder until it
vanishes in the umbra during the transit. When the source is seen
again, the inverse behaviour is observed.\\

\section{Chromaticity effects in extragalactic microlensing}

Let us now turn to the extragalactic microlensing case, which was
previously studied by Torres et al. \cite{torr} using point
sources. We shall study the lightcurves produced by a stellar-mass
extragalactic wormhole when it crosses the line of sight to a
background compact source, namely and Active Galactic Nuclei
(AGN). As discussed by Torres et al. \cite{torr}, a critical
requirement to observe a significant light magnification during
the microlensing event is that the size of the source, when
projected onto the lens plane, is not larger than the Einstein
ring of the lensing mass. This is necessary because otherwise
light from outside the ring could become dominant and smooth out
the gravitationally induced variability. In the case of an AGN,
since the size of the emitting regions varies with the wavelength,
the gravitational amplification will be more effective at those
wavelengths at which the source is more compact. The wavelengths
where the amplifications are maximized are located at gamma-ray
energies \cite{torr}.\\

The compactness of the gamma-ray central region of an AGN is
determined by the opacity to the propagation of gamma-ray photons
in the X-ray field produced by the object. Since the opacity is a
function of the photon energy, the size of the observable
gamma-ray photosphere will also depend on the observing energy.
Differential amplification during the microlensing event, then,
should lead to chromaticity effects. In order to compute these
effects, we shall adopt a radius of the photosphere given by the
maximum height at which photons of energy $E$ are absorbed by pair
creation in the X-ray radiation field of the inner accretion disk.
According to Becker \& Kafatos \cite{becker}, this size is
\begin{equation}
r_{\gamma }\propto E^{\alpha/(2\alpha +3)},  \label{e1}
\end{equation}
where $\alpha$ is the X-ray spectral index of the accretion disk
radiation field. In our calculations, we shall adopt an average
value $\alpha=1.1$ \cite{kro}. The larger photospheres, then, are
those observed at the higher energies. To compute the microlensing
effects on the AGN, we define a reference source with radius
$r_{\rm ref}$ and gamma-ray energy $E_{\rm ref}$, such that
\begin{equation}
R_{\gamma }(E)=R_{\rm ref}\left( \frac{E}{E_{\rm ref}}\right)
^{\alpha/(2\alpha +3)},  \label{e3}
\end{equation}
where we have written $r_\gamma$ and $r_{\rm ref}$ in units of the
Einstein radius $R_{\rm E}$.\\

The intensity of the source (without lensing) is uniform and its
spectrum follows approximately a power law:
\begin{equation}
I_{0}(E)\propto E^{-\xi }  \label{e4}
\end{equation}
with $\xi$ typically in the range $\xi \in (1.7,2.7)$ . We can
write this latter Eq.  in the form:
\begin{equation}
I_{0}(E)=I_{\rm ref}\left( \frac{E}{E_{\rm ref}}\right) ^{-\xi },
\label{e5}
\end{equation}
where $I_{{\rm ref}}$ is the intensity of the reference source.
The surface intensity distribution is assumed to be constant at a
given energy
\begin{equation}
{\mathcal I}_{0}(E)=\frac{I_{0}(E)}{\pi \left( R_{\gamma
}(E)\right) ^{2}}. \label{e6}
\end{equation}
Then, the total amplification of the source due to microlensing
can be computed as
\begin{equation}
A=\frac{I}{I_{0}}=\frac{\int_{0}^{2\pi }\int_{0}^{R_{\gamma }(E)}{\mathcal I}%
_{0}(E)A_{0}(R,\varphi )RdRd\varphi }{\int_{0}^{2\pi
}\int_{0}^{R_{\gamma}(E)}{\mathcal I}_{0}(E)RdRd\varphi }.
\label{e7}
\end{equation}
Since ${\mathcal I}_{0}(E)$ does not depend on $(R,\varphi )$, we
have
\begin{equation}
A=\frac{\int_{0}^{2\pi }\int_{0}^{R_{\gamma }(E)}A_{0}(R,\varphi
)RdRd\varphi }{\pi \left( R_{\gamma }(E)\right) ^{2}}.  \label{e8}
\end{equation}
\mbox{}From Eq.  (\ref{e5}) and Eq.  (\ref{e7}) we obtain:
\begin{equation}
I=AI_{0}=AI_{\rm ref}\left( \frac{E}{E_{\rm ref}}\right) ^{-\xi
}.\label{e9}
\end{equation}
Let us now introduce the dimensionless minimum impact parameter
$k$ as:
\begin{equation}
k=\frac{B_{0}}{R_{\gamma}(E)}. \label{e11}
\end{equation}
Adopting $\xi =2$, $R_{\rm ref}=0.2$, and $E_{\rm ref}=100$ MeV,
we can use Eq.  (\ref{e8}) and Eq.  (\ref{e9}) to compute the
expected lightcurves and the spectral evolution during the
microlensing event. The results are shown, for a variety of impact
parameters, in Figures 4-7, where
we define $J$ as the ratio between $I$ and $I_{\rm ref}$. \\

The main point to be emphasized in the spectral evolution of the
lensed source is that, when the source is a wormhole, the spectrum
emerges from the umbra at its lowest level (the bottom curve in
the spectral plots) and then it jumps to its maximum, from where
it decreases slowly keeping its slope in the log-log diagram. This
occurs because after the transit the intensity have to recover
from the umbra null values. The effects of the differential
obscuration also appear as a clear break in the spectrum at early
times since the transit. This break, with a hardening at low
energies, results because the smaller and less energetic
gamma-spheres emerge from the umbra after than the more energetic
(and bigger) ones, and then are initially less amplified. The
break in the spectrum is best appreciated in Figure 6, where the
impact parameter is $k=1$.\\

In the case of positive lenses (upper set of panels in the
figures) the evolution is smoother, decreasing the emission at all
wavelengths after the transit. The chromaticity effects in this
case are limited to a small slope change at medium energies.
\begin{figure}[t]
\begin{center}
\includegraphics[width=8cm,height=7cm]{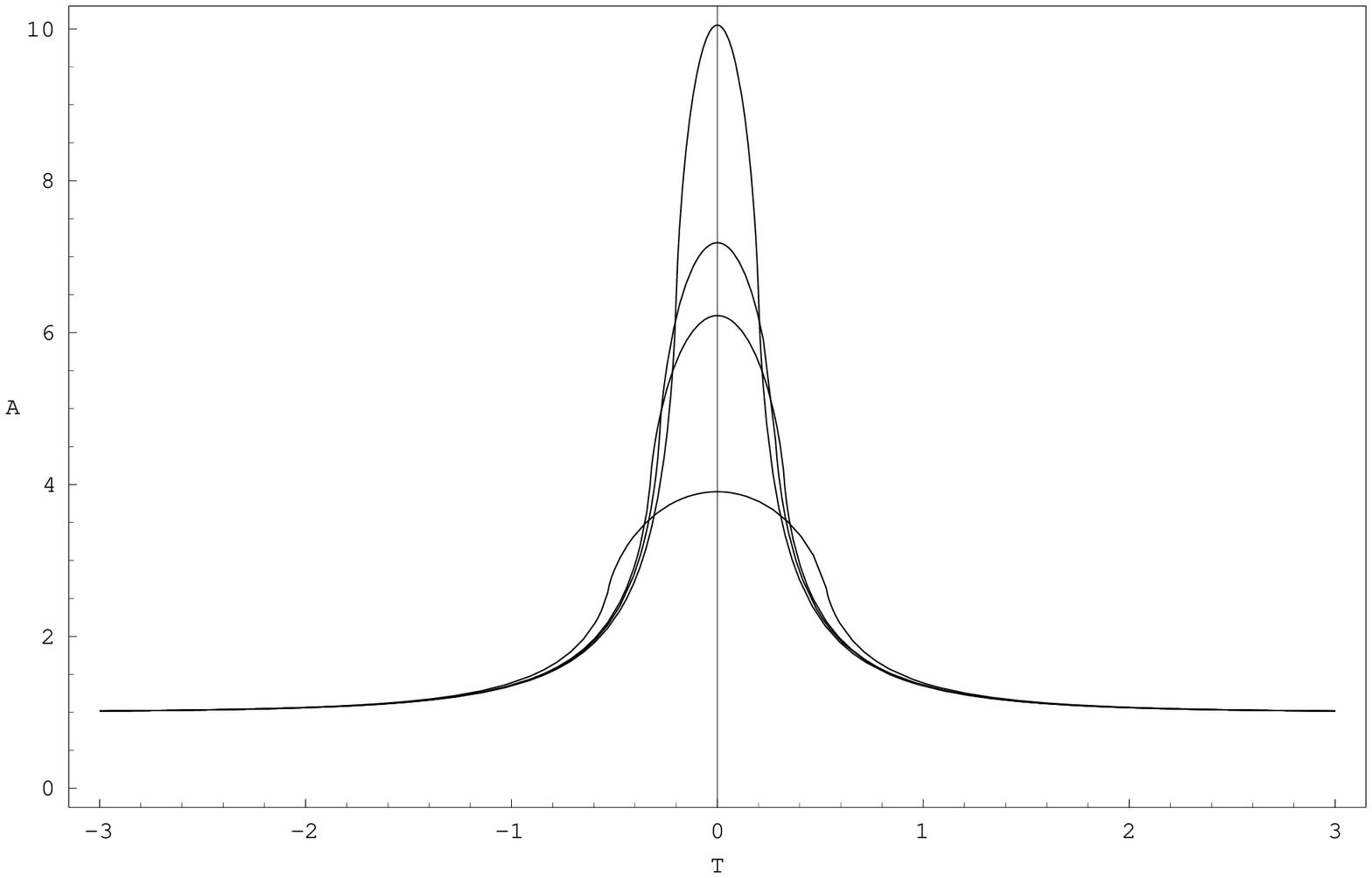}%
\includegraphics[width=8cm,height=7cm]{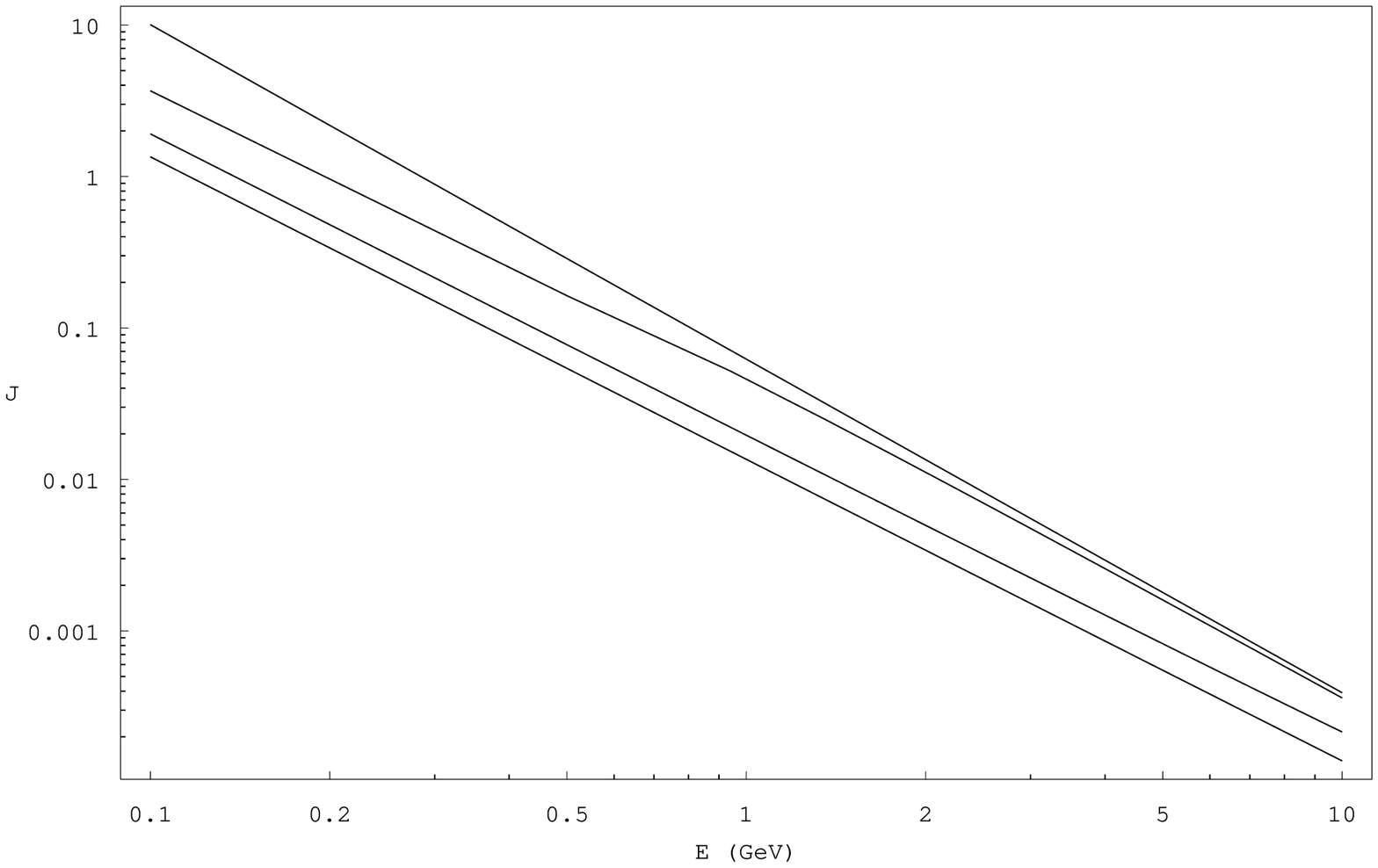}
\includegraphics[width=8cm,height=7cm]{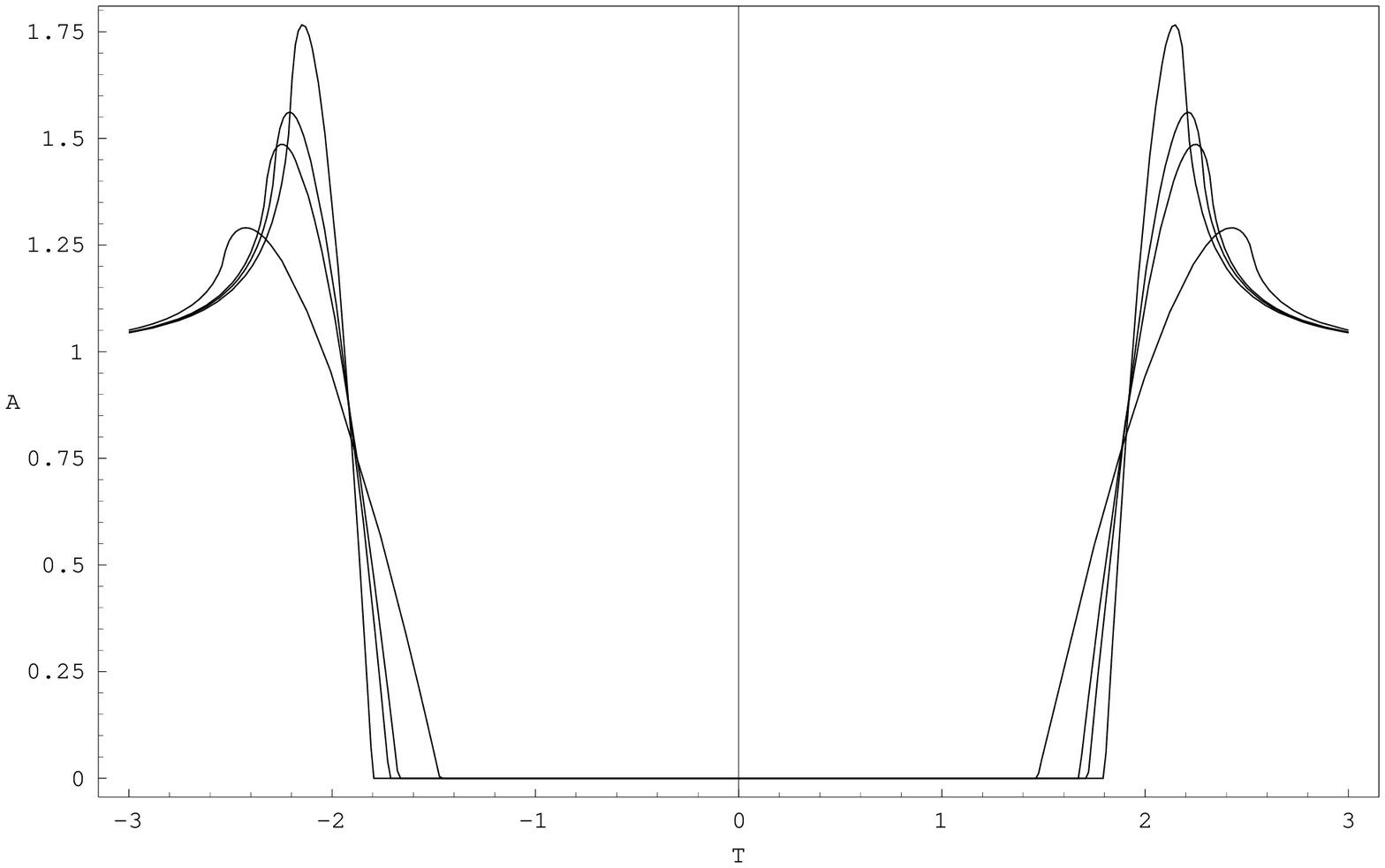}
\includegraphics[width=8cm,height=7cm]{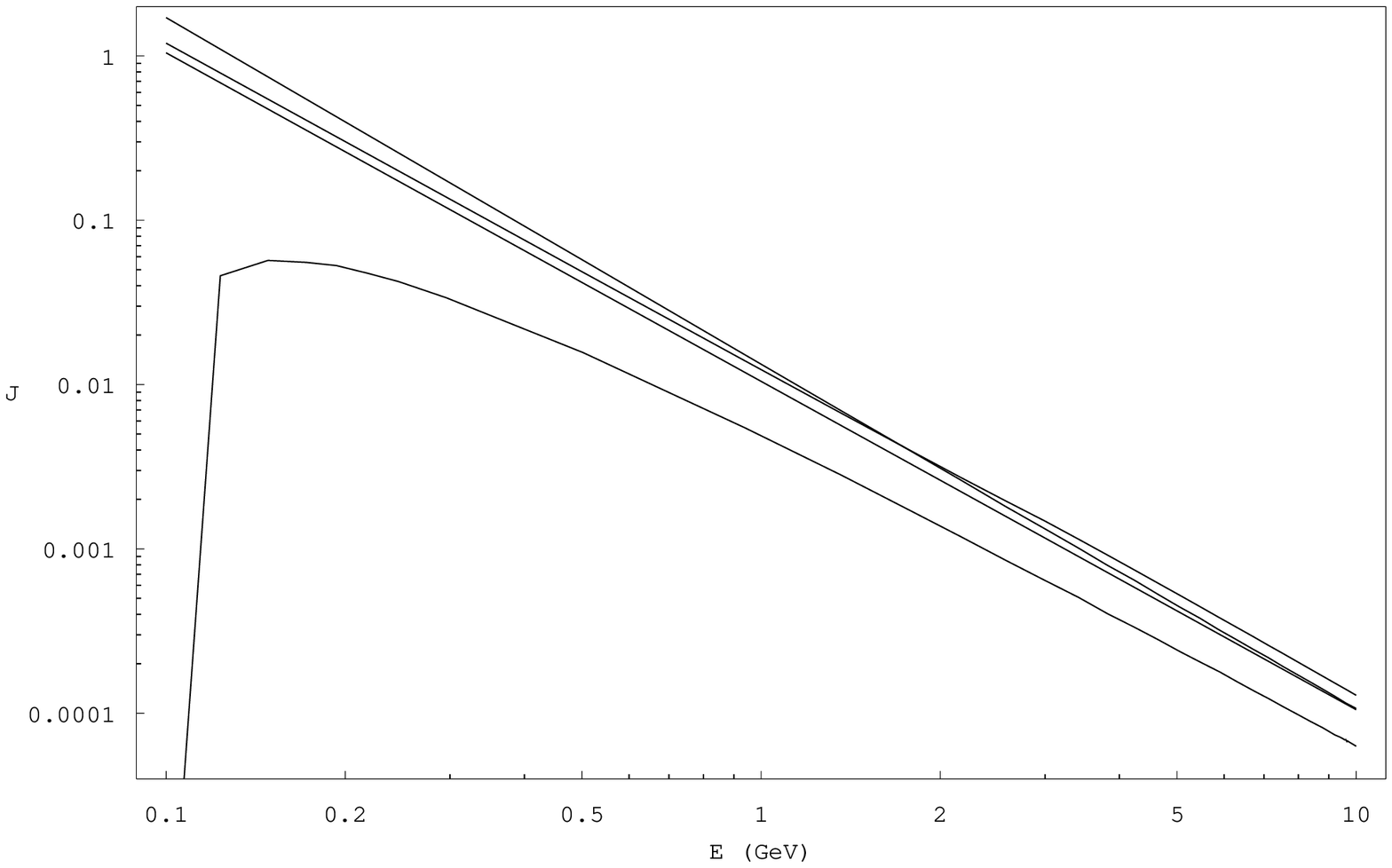}
\end{center}
\vspace{-3cm} \caption{Lightcurves (on the left) for a
microlensing event of a background gamma-ray emitting AGN produced
by a positive point mass (upper panel) and a negative point mass
(lower panel). Impact parameter $k=0$. The different curves
correspond to the amplification of different regions in the
object. The energies considered were 10GeV, 1GeV, 500MeV, and
100MeV, and correspond, both for positive and negative lensing, to
the curves depicted on the left diagrams from bottom to top. The
corresponding spectral changes for the entire source are shown in
the panels on the right side. For the positive case the spectra
evolve from the top to downwards as time increases from $T=0$ to
$T=1$. In the negative case, instead, the evolution of the curves
with time is given by, from top downwards, $T=2.1$, $T=2.4$,
$T=3.0$, and finally the curve for $T=1.8$. See text for further
explanation. \label{fig3}} \end{figure}
\begin{figure}[t]
\begin{center}
\includegraphics[width=8cm,height=7cm]{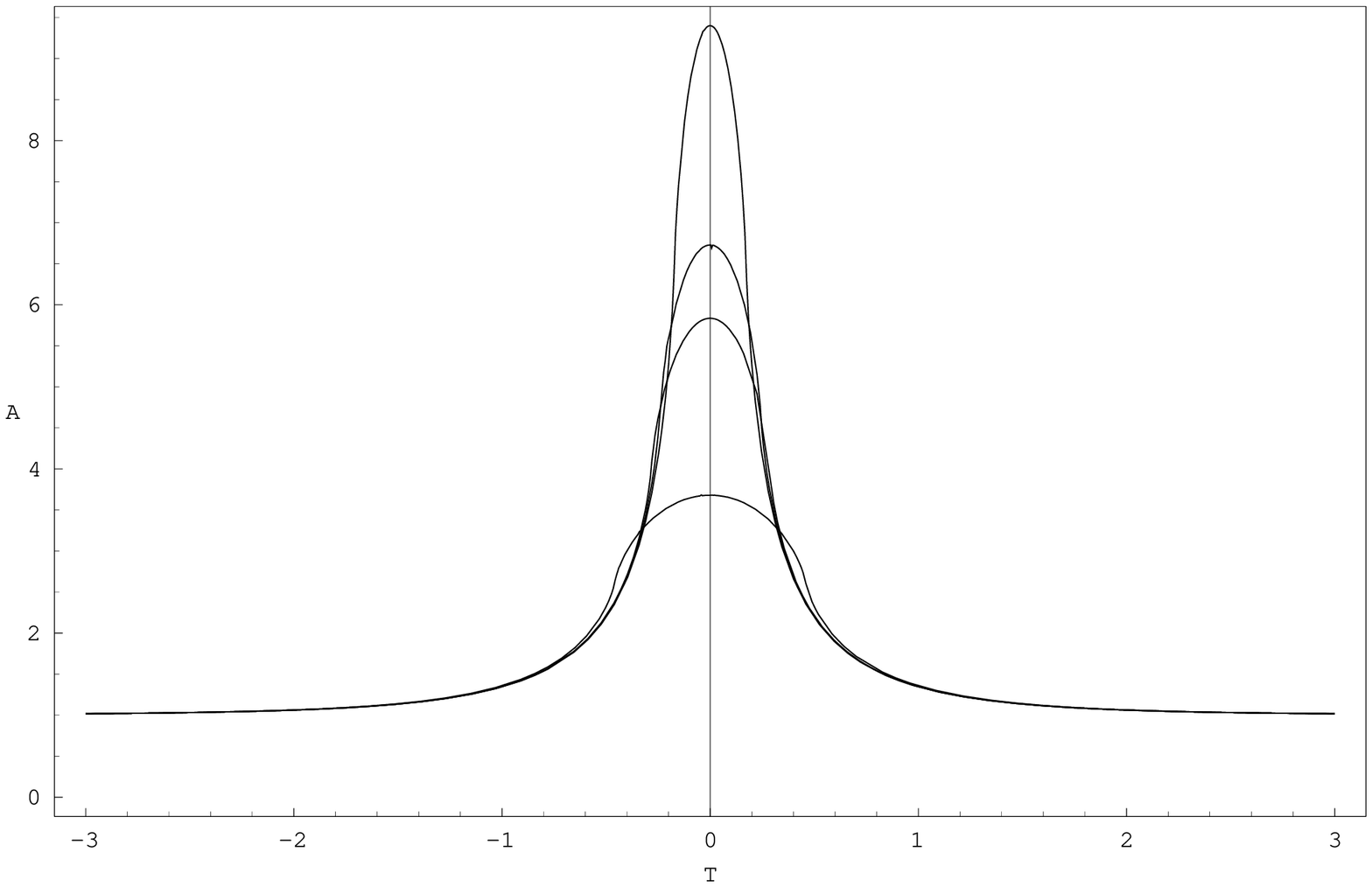}%
\includegraphics[width=8cm,height=7cm]{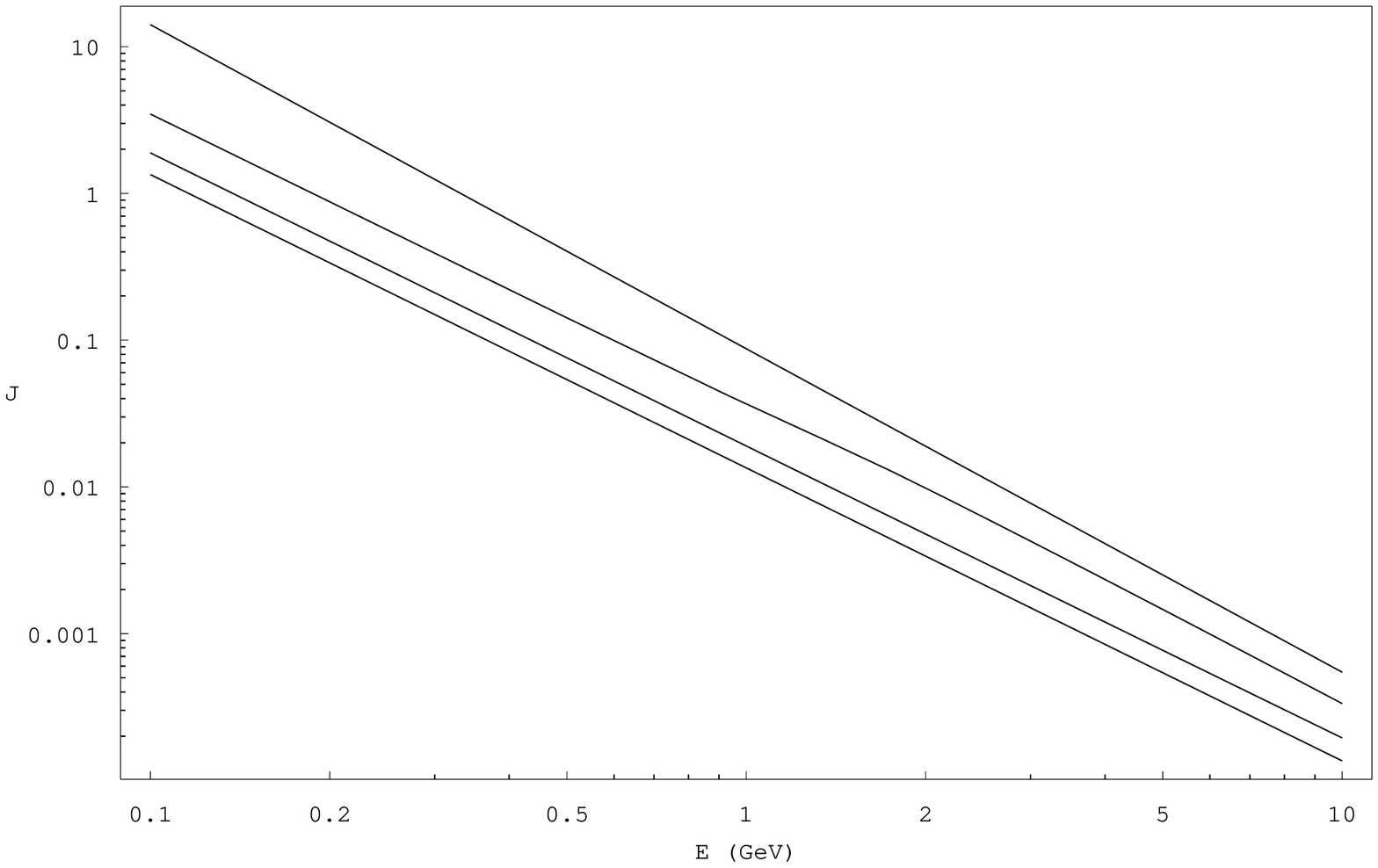}
\includegraphics[width=8cm,height=7cm]{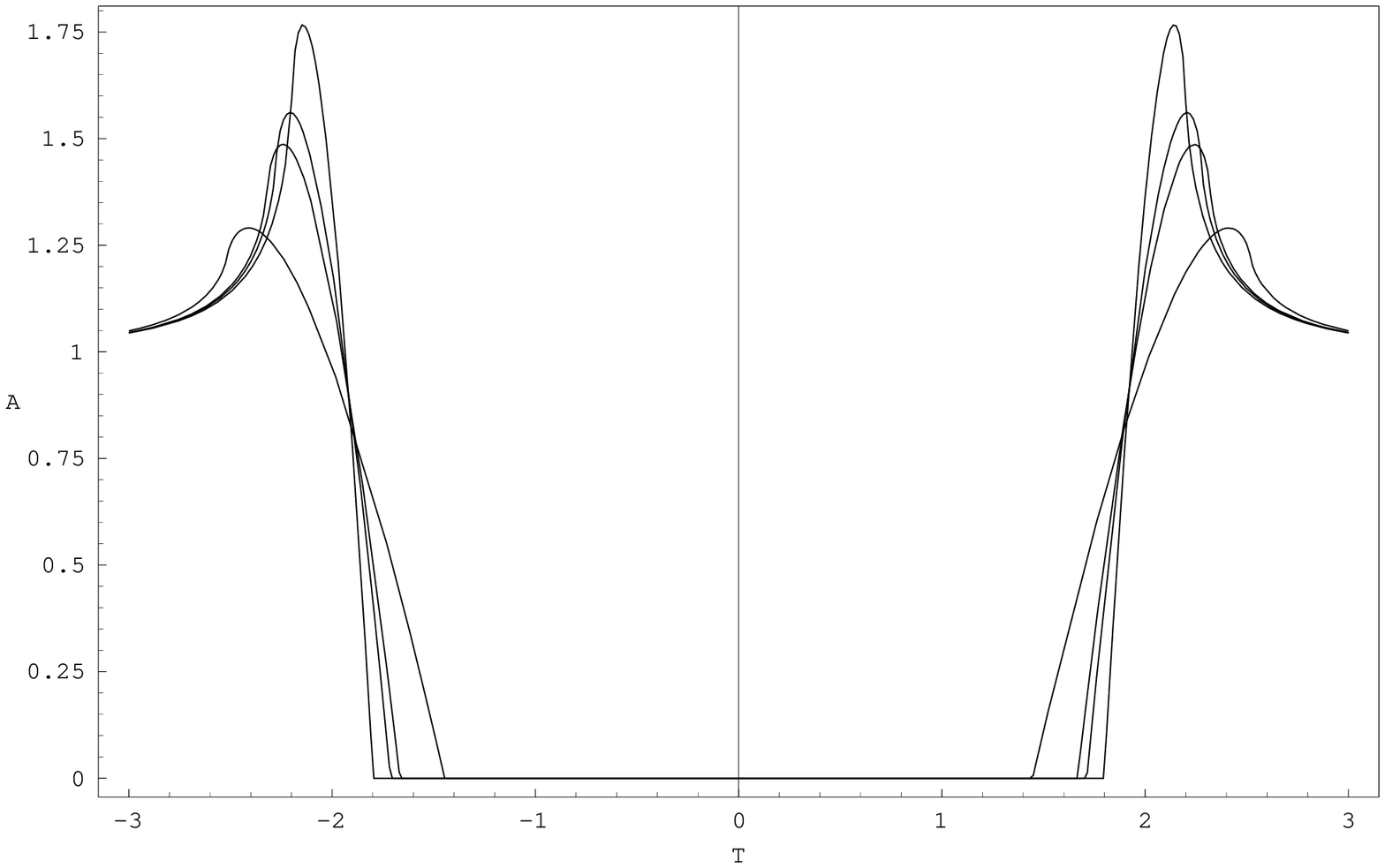}
\includegraphics[width=8cm,height=7cm]{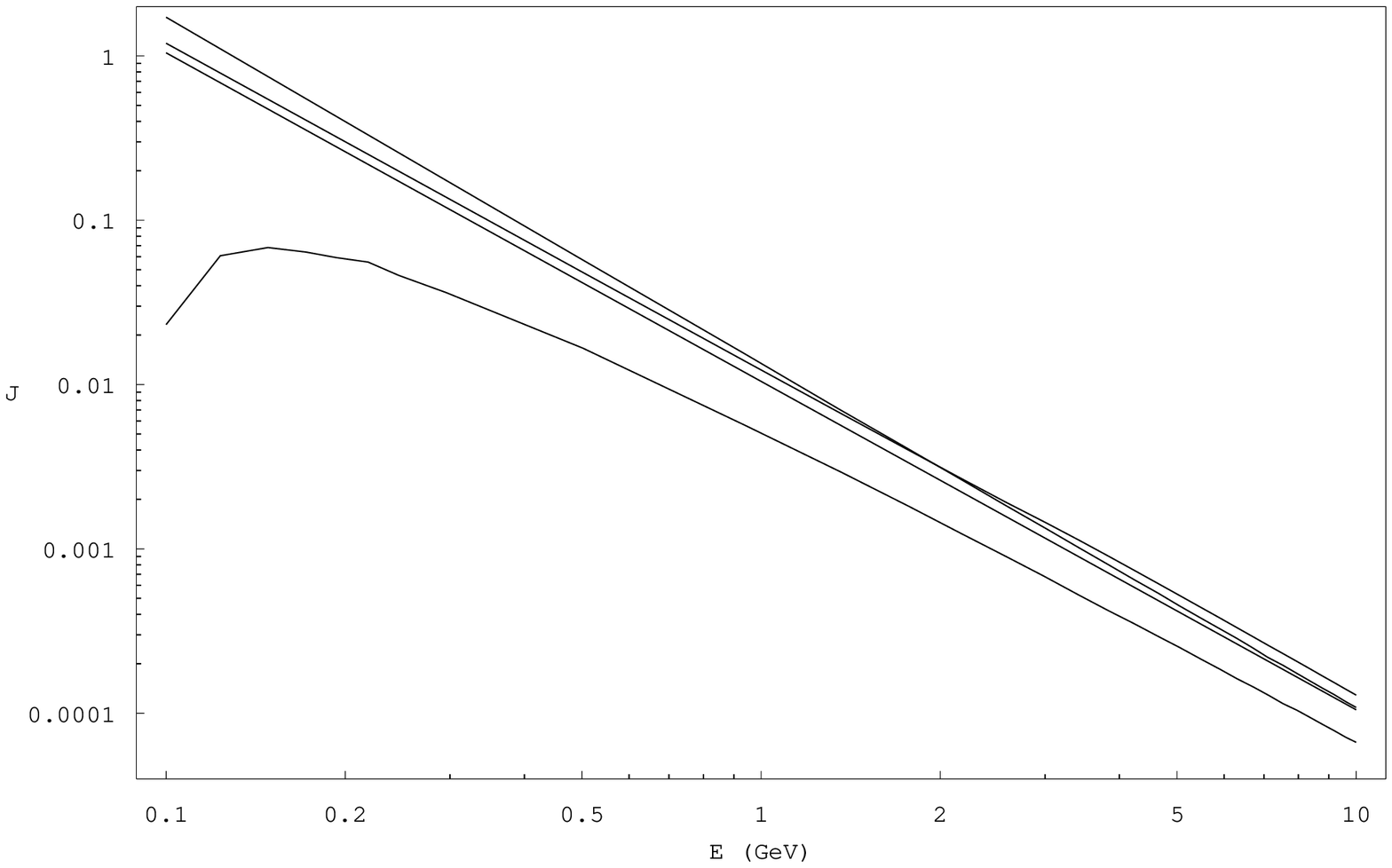}
\end{center}
\vspace{-3cm} \caption{Idem Fig. 4, but for $k=0.5$ \label{fig4}}
\end{figure}

\begin{figure}[t]
\begin{center}
\includegraphics[width=8cm,height=7cm]{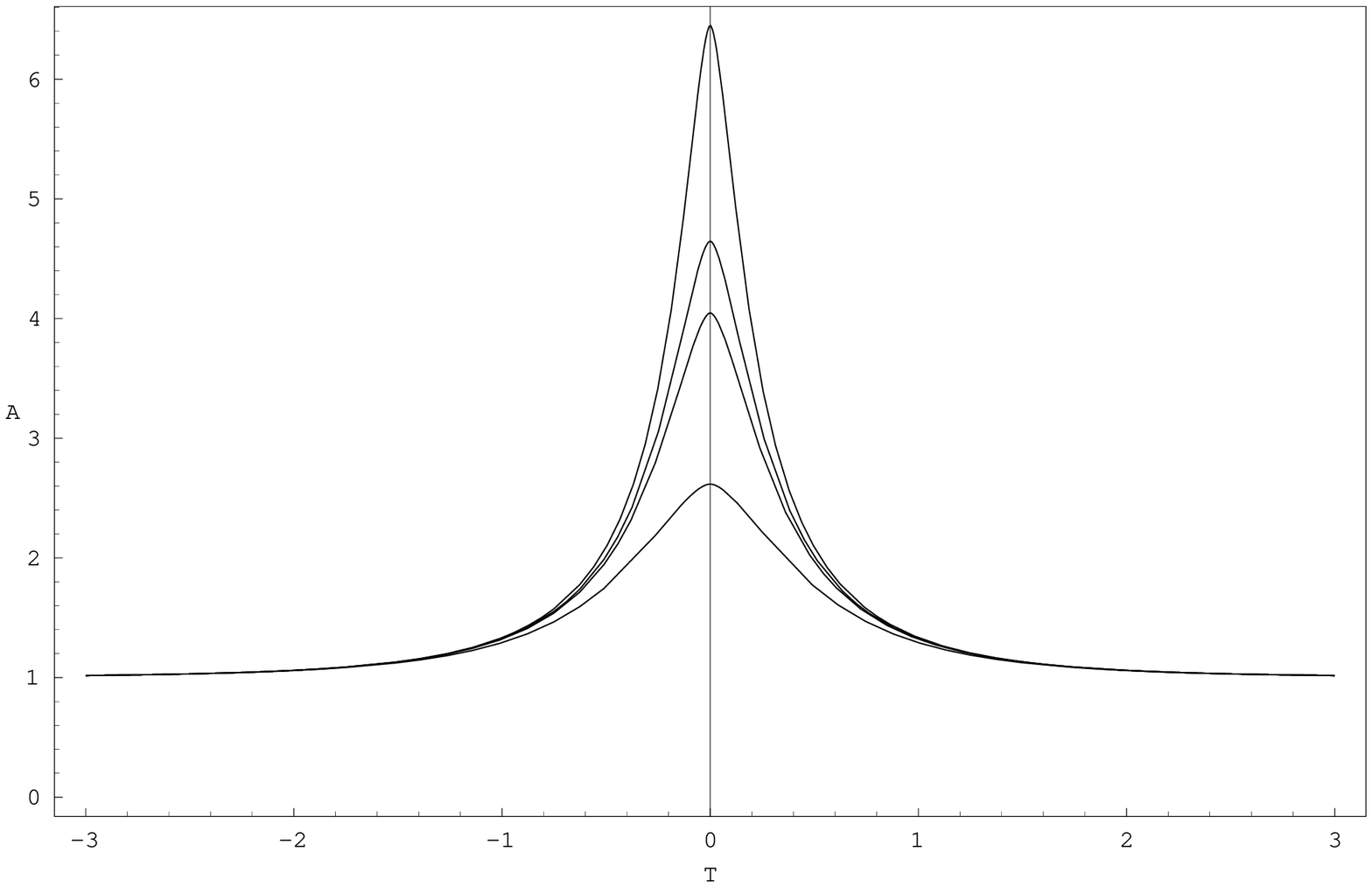}%
\includegraphics[width=8cm,height=7cm]{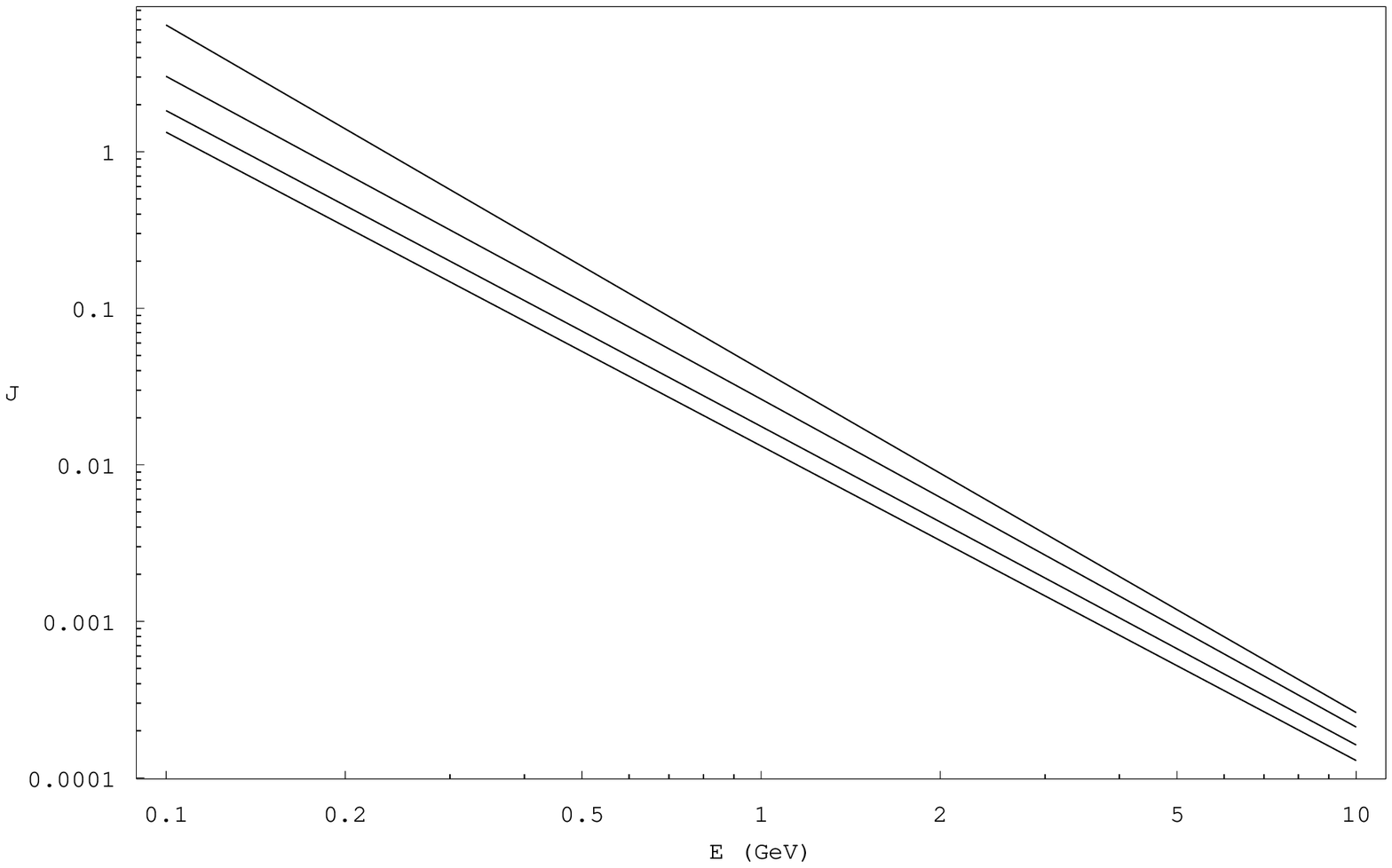}
\includegraphics[width=8cm,height=7cm]{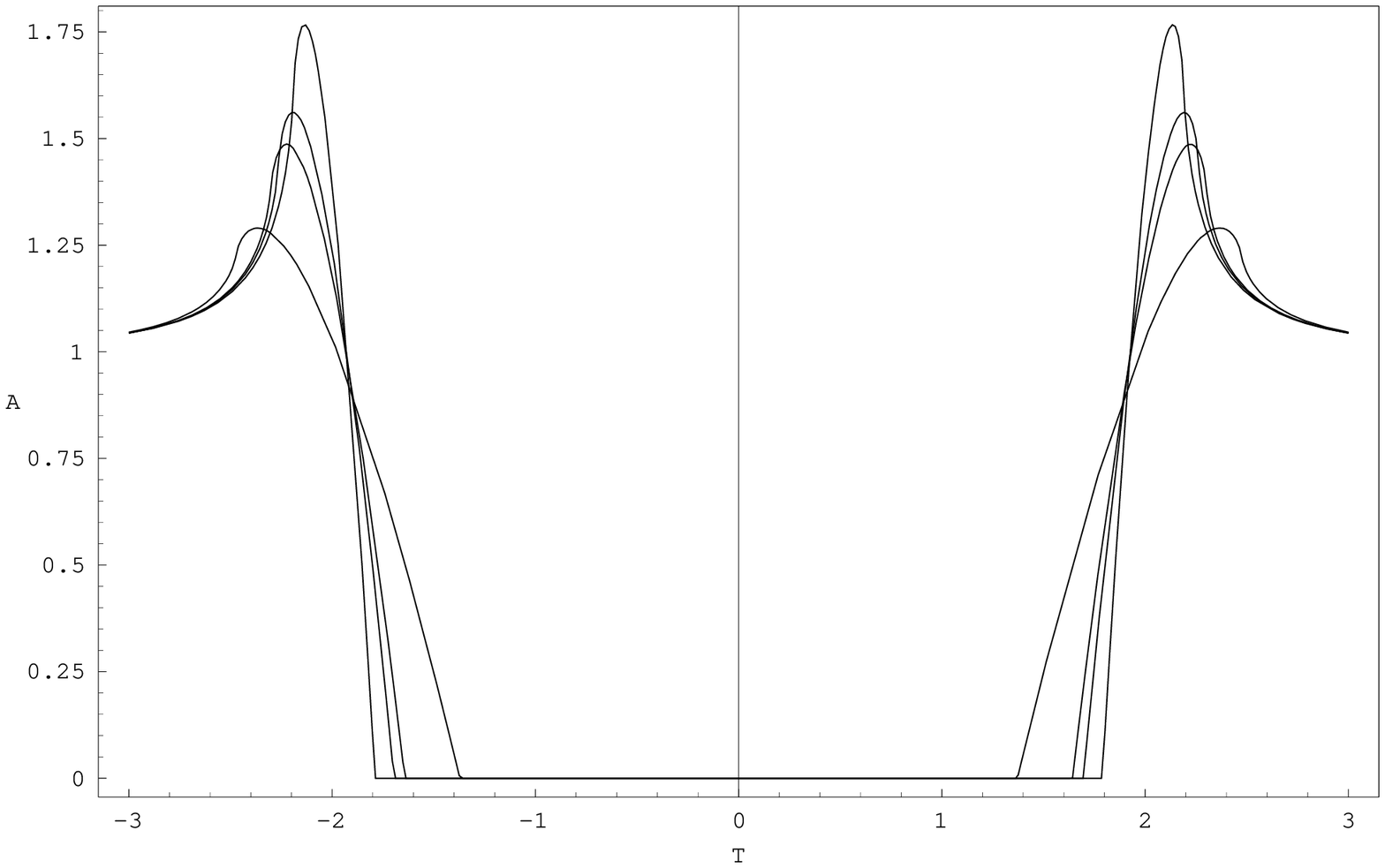}
\includegraphics[width=8cm,height=7cm]{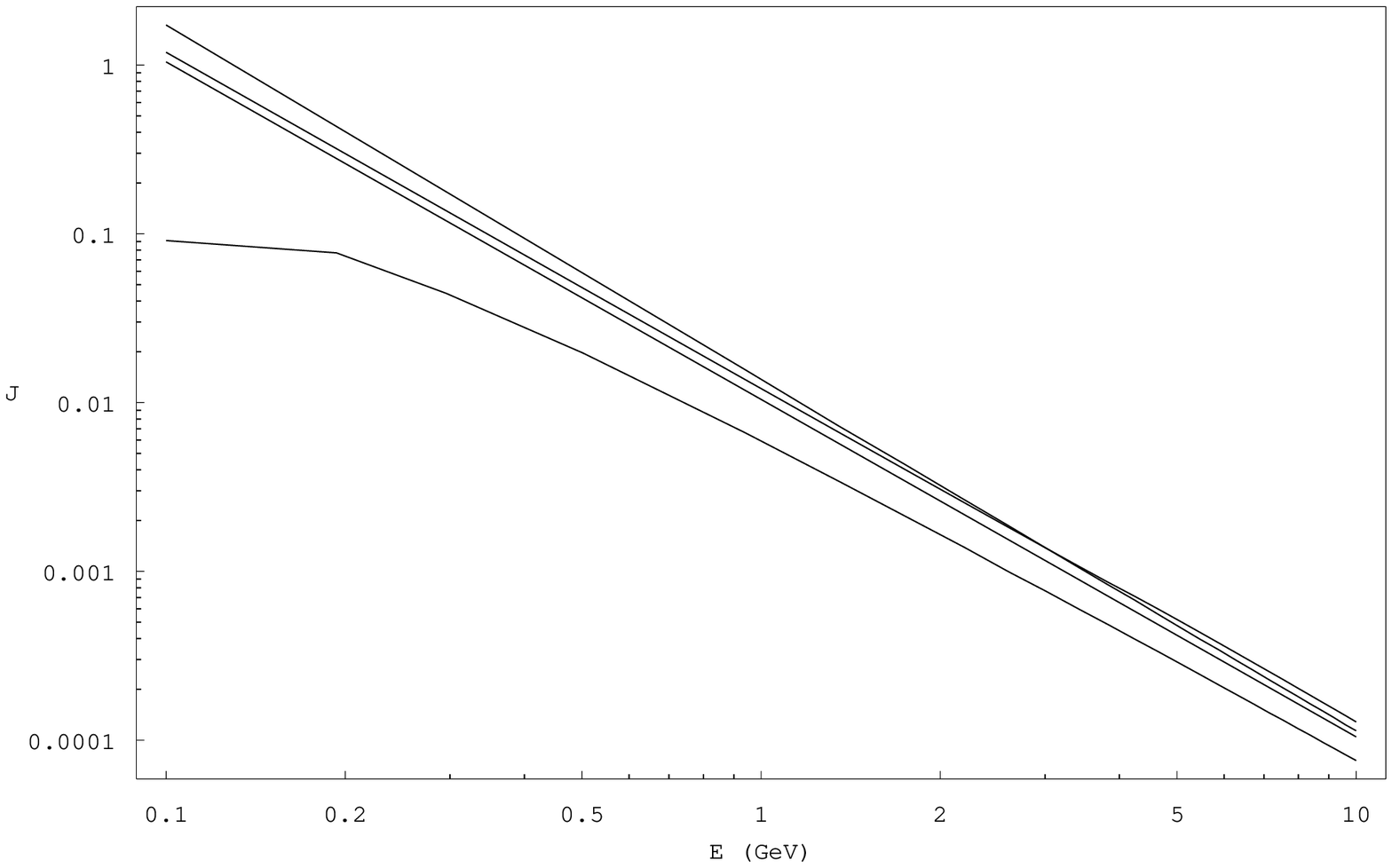}
\end{center}
\vspace{-3cm} \caption{Idem Fig. 4, but for $k=1$ \label{fig5}}
\end{figure}
\begin{figure}[t]
\begin{center}
\includegraphics[width=8cm,height=7cm]{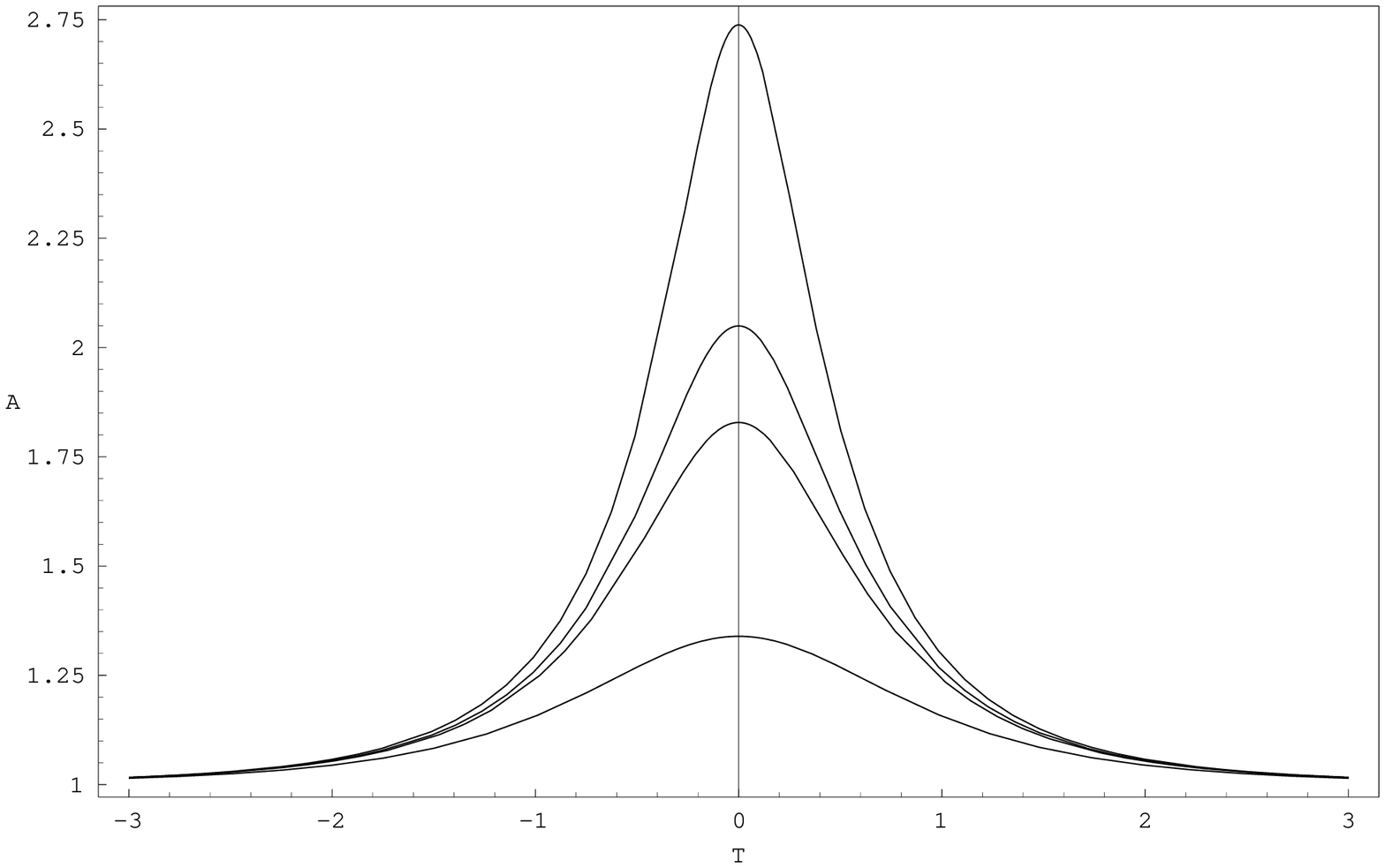}%
\includegraphics[width=8cm,height=7cm]{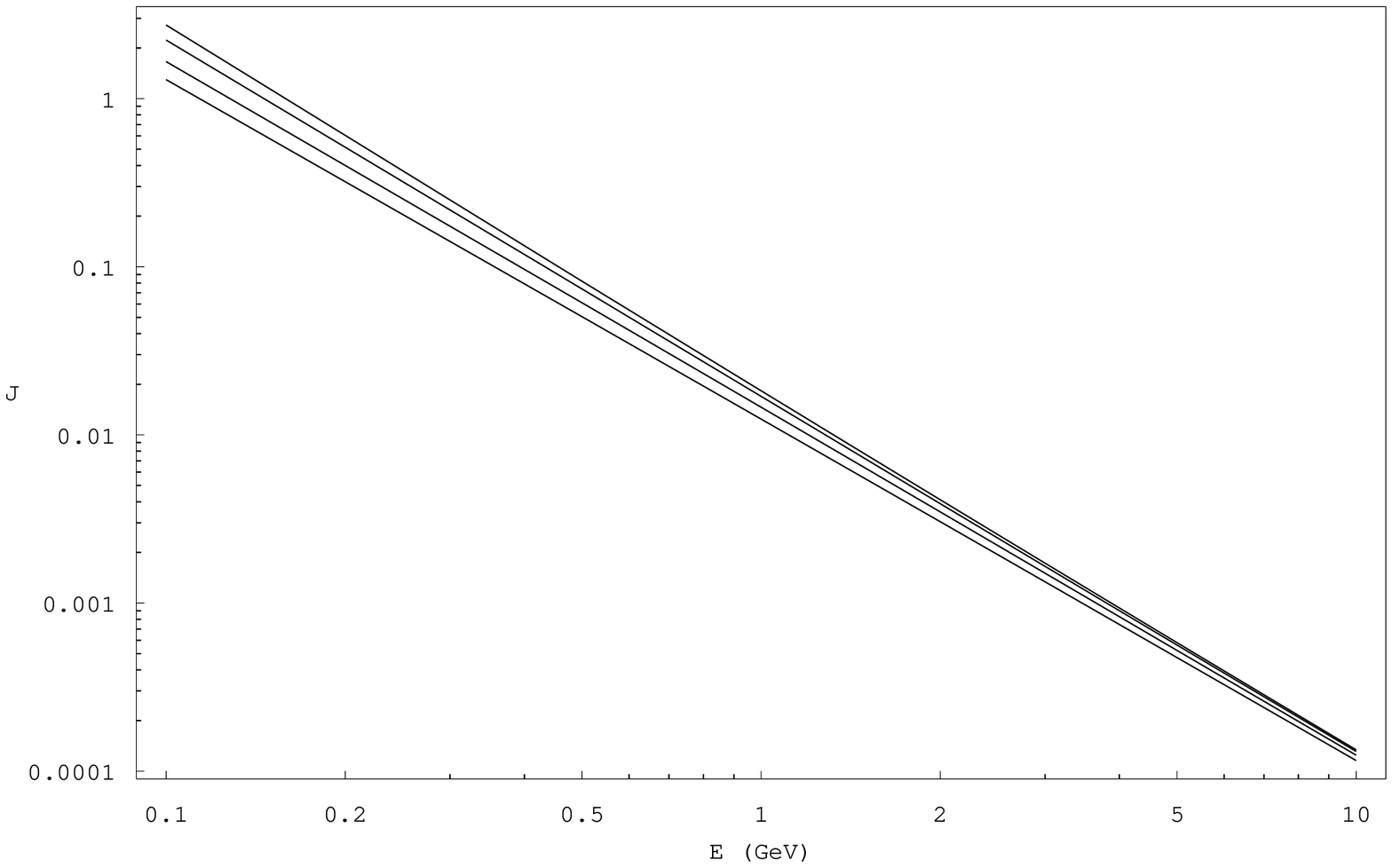}
\includegraphics[width=8cm,height=7cm]{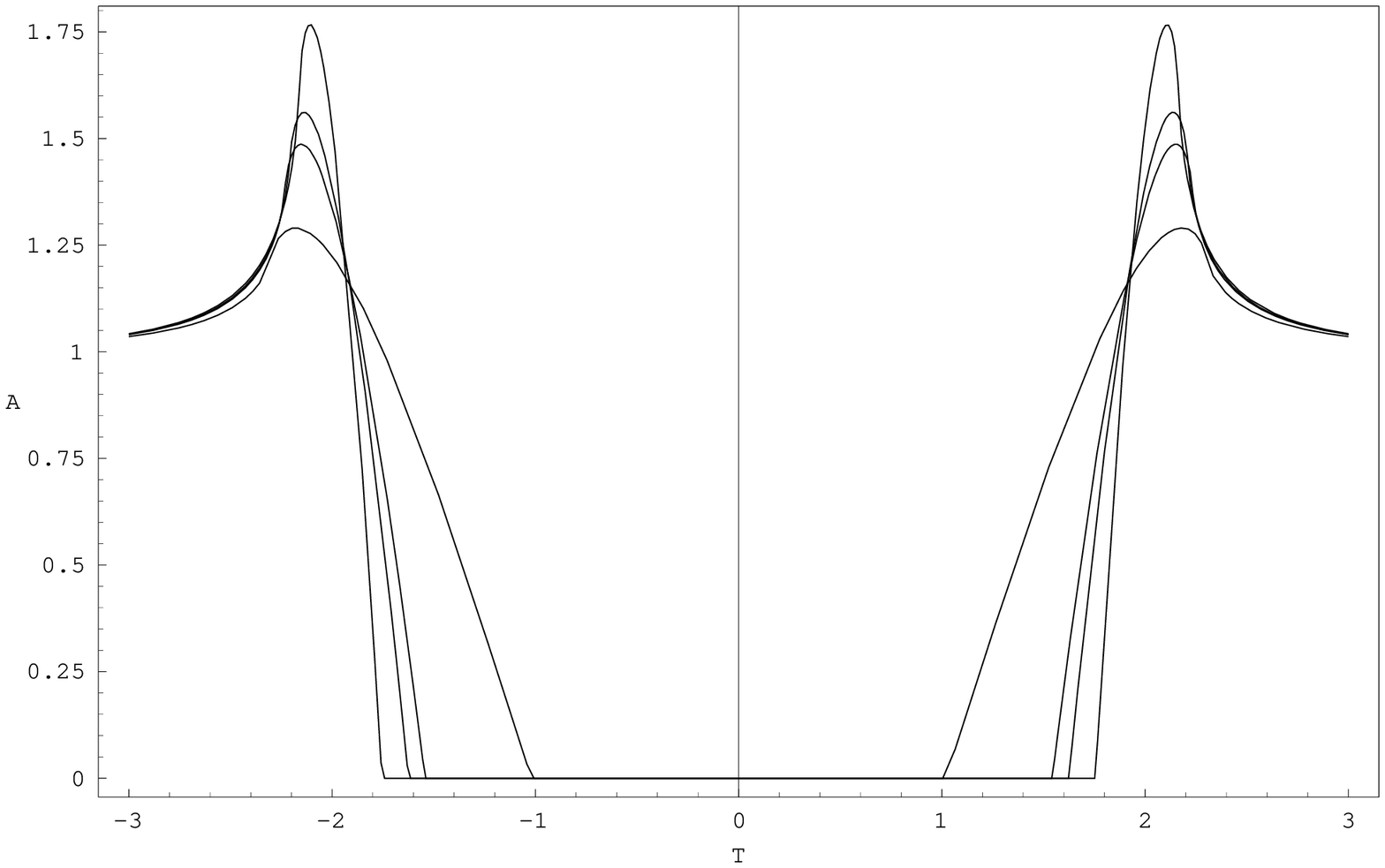}
\includegraphics[width=8cm,height=7cm]{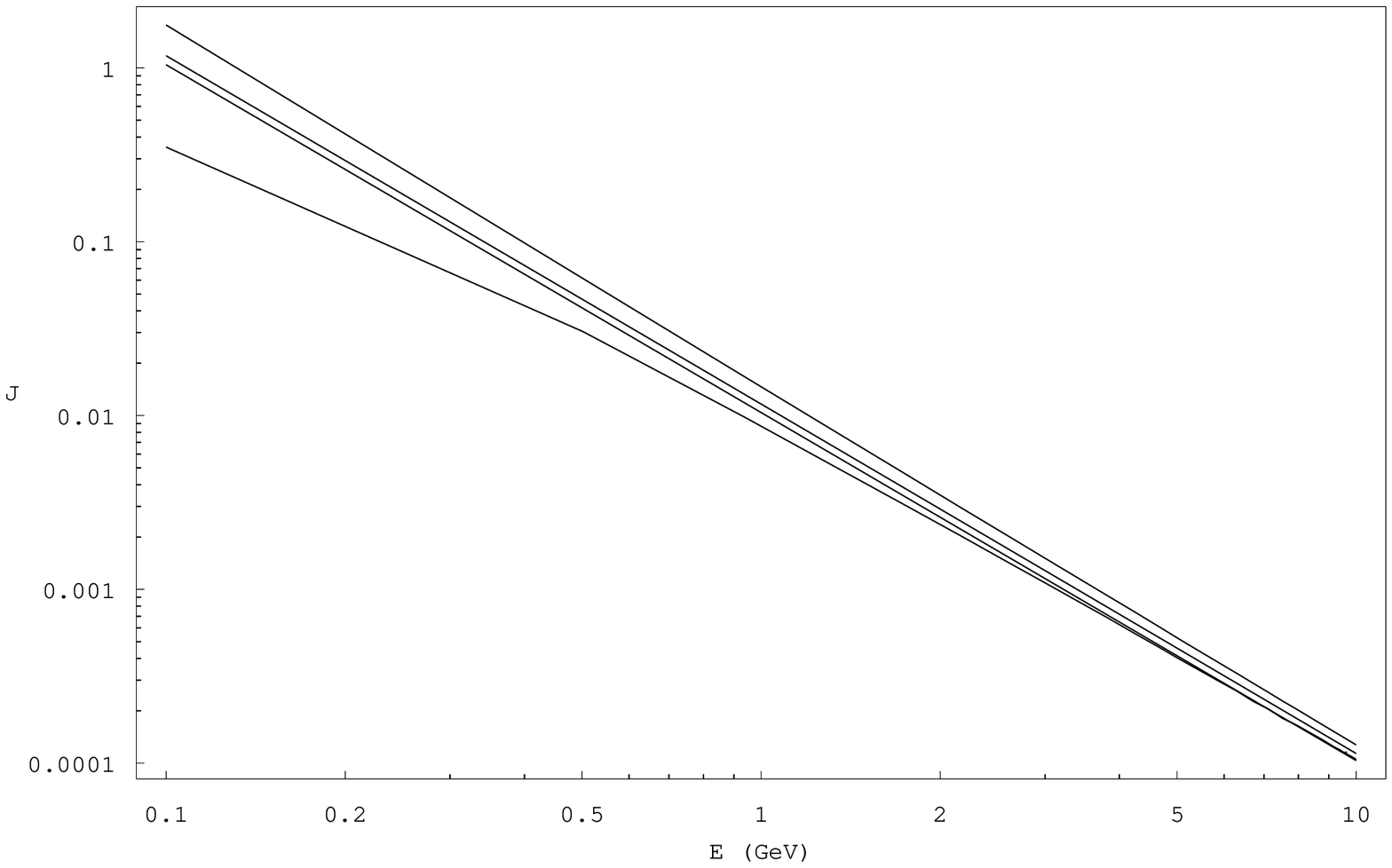}
\end{center}
\vspace{-3cm} \caption{Idem Fig. 4, but for $k=2$ \label{fig6}}
\end{figure}

\section{Concluding remarks}

In this paper we have developed the formalism for gravitational
microlensing of extended sources when the lens is a wormhole-like
object. We have studied the chromaticity effects introduced by the
finite source size and the non-uniform intensity distribution on
it, finding a set of very peculiar features that allow a clear
identification of any lens having a negative energy density. These
chromaticity effects, when combined with the observed lightcurves,
provide a unique tool to study the possible existence of natural
wormholes in the universe.

\section*{Acknowledgments}
This work has been supported by Universidad de Buenos Aires
(UBACYT 01TW77, EE), CONICET (DFT, and PIP 0430/98, GER), ANPCT
(PICT 98 No. 03-04881, GER), and Fundaci\'{o}n Antorchas (GER and
DFT).




\begin{thebibliography}{99}



\bibitem{motho}M. S. Morris \& K. S. Thorne, Am. J. Phys. {\bf 56}, 395 (1988).

\bibitem{wh}D. Hochberg \& M. Visser, Phys. Rev. Lett. {\bf 81},
746 (1998); Phys. Rev D{\bf 58}, 044021 (1998); Phys. Rev. D{\bf
56}, 4745 (1997); E. E. Flanagan \& R. M. Wald, Phys. Rev. D{\bf
54}, 6233 (1996); L. A. Anchordoqui, S. E. Perez Bergliaffa \& D.
F. Torres, Phys. Rev. D{\bf 55}, 5226 (1997); C. Barcel\'o \& M.
Visser, Phys. Lett. B{\bf 466}, 127 (1999).

\bibitem{cramer} J. G. Cramer, R. L. Forward, M. S. Morris, M. Visser, G. Benford, G. A.
Landis, Phys. Rev. D{\bf 51}, 3117 (1995).

\bibitem{torr}D. F. Torres, G. E. Romero \& L. A. Anchordoqui,
Phys. Rev. D{\bf 58}, 123001 (1998); D. F. Torres, G. E. Romero \&
L. A. Anchordoqui, {\it (Honorable Mention, Gravity Foundation
Research Awards 1998)}, Mod. Phys. Lett. {\bf A13}, 1575 (1998).

\bibitem{S} M. Safonova, G. E. Romero \& D. F. Torres,
Mod. Phys. Lett. {\bf A16}, 153 (2001) [astro-ph/0104075].

\bibitem{doqui}L. A. Anchordoqui, G. E. Romero, D.
F. Torres \& I. Andruchow, Mod. Phys. Lett. {\bf A14}, 791 (1999).

\bibitem{rom}G.E. Romero, D.F. Torres, L.A. Anchordoqui, I. Adruchow,
B. Link, Monthly Notices Royal Astron. Soc. {\bf 308}, 799 (1999).

\bibitem{Zwart}S. F. Portegies Zwart, C-H. Lee, H.K. Lee, Astrophys. J.
{\bf 520}, 666 (1999).

\bibitem{bin} A. Udalski, M. Szyma\'{n}ski, S. Mao, et al.,
Astrophys. J. {\bf 436}, L103 (1994).

\bibitem{becker}P.A. Becker, M. Kafatos, Astrophys. J. {\bf 453}, 83
(1995).

\bibitem{Han}C. Han, S-H. Park, J-H. Jeong, Monthly Notices Royal Astron. Soc.
{\bf 316}, 97 (2000).

\bibitem{kro}J.H. Krolik, {\it Active Galactic Nuclei}, (Princeton
University Press, Princeton, 1999).








\end{thebibliography}
\end{document}